\documentclass[12pt]{article}

\usepackage{amsfonts}
\usepackage[centertags]{amsmath}
\usepackage{amssymb}
\usepackage{cite}
\usepackage{psfrag}
\usepackage{graphicx}
\usepackage{subfig}
\usepackage{bbm,bm}
\usepackage{epsfig, multicol}

\allowdisplaybreaks
\begin{document}

\topmargin 0pt
\oddsidemargin 0mm
\def\be{\begin{equation}}
\def\ee{\end{equation}}
\def\bea{\begin{eqnarray}}
\def\eea{\end{eqnarray}}
\def\ba{\begin{array}}
\def\ea{\end{array}}
\def\ben{\begin{enumerate}}
\def\een{\end{enumerate}}
\def\nab{\bigtriangledown}
\def\tpi{\tilde\Phi}
\def\nnu{\nonumber}
\newcommand{\eqn}[1]{(\ref{#1})}

\newcommand{\half}{{\frac{1}{2}}}
\newcommand{\vs}[1]{\vspace{#1 mm}}
\newcommand{\dsl}{\pa \kern-0.5em /} 
\def\a{\alpha}
\def\b{\beta}
\def\g{\gamma}\def\G{\Gamma}
\def\d{\delta}\def\D{\Delta}
\def\ep{\epsilon}
\def\et{\eta}
\def\z{\zeta}
\def\t{\theta}\def\T{\Theta}
\def\l{\lambda}\def\L{\Lambda}
\def\m{\mu}
\def\f{\phi}\def\F{\Phi}
\def\n{\nu}
\def\p{\psi}\def\P{\Psi}
\def\r{\rho}
\def\s{\sigma}\def\S{\Sigma}
\def\ta{\tau}
\def\x{\chi}
\def\o{\omega}\def\O{\Omega}
\def\k{\kappa}
\def\pa {\partial}
\def\ov{\over}
\def\nn{\nonumber\\}
\def\ud{\underline}
\begin{flushright}
%
\end{flushright}
\begin{center}
{\large{\bf Anisotropic SD2 brane: accelerating cosmology\\ and      
Kasner-like space-time from compactification}}

\vs{10}

{Kuntal Nayek\footnote{E-mail: kuntal.nayek@saha.ac.in} and Shibaji Roy\footnote{E-mail: shibaji.roy@saha.ac.in}}

\vs{4}

{\it Saha Institute of Nuclear Physics\\
1/AF Bidhannagar, Calcutta 700064, India\\}

\vs{4}

and

\vs{4}

{\it Homi Bhabha National Institute\\
Training School Complex, Anushakti Nagar, Mumbai 400085, India}

\end{center}

\vs{15}

\begin{abstract}
Starting from an anisotropic (in all directions including the time direction of the brane) non-susy D2 brane solution of type IIA string theory
we construct an anisotropic space-like D2 brane (or SD2 brane, for short) solution by the standard trick of double Wick rotation. This solution
is characterized by five independent parameters. We show that compactification on six dimensional hyperbolic space (H$_6$) of time dependent
volume of this SD2 brane solution leads to accelerating cosmologies (for some time $t\sim\,t_0$, with $t_0$ some characteristic time) where 
both the expansions and the accelerations are 
different in three spatial directions of the resultant four dimensional universe. On the other hand at early times ($t \ll t_0$)
this four dimensional space, in certain situations, leads to four dimensional Kasner-like cosmology, with two additional scalars, namely, the dilaton 
and a volume scalar of H$_6$. 
Unlike in the standard four dimensional Kasner cosmology here all three Kasner exponents could be positive definite, leading to expansions in
all three directions.       
\end{abstract}
\newpage

\noindent{\it 1. Introduction} : It is well-known \cite{Townsend:2003fx} that cosmological solution of higher dimensional vacuum Einstein 
equation can give rise to interesting 
four dimensional cosmology (with a period of accelerated expansion) upon time dependent hyperbolic space compactifications \cite{Kaloper:2000jb}. 
This process, therefore,
evades a no-go theorem \cite{Gibbons:1985,Maldacena:2000mw} of obtaining such accelerated expansion in standard time-independent compactifications. 
Similar cosmologies also follow
if one includes fluxes \cite{Ohta:2003pu} and/or a dilaton field \cite{Roy:2003nd} in the higher dimensional theories such as M/string theory. 
M/string theory solution which gives rise
to four dimensional accelerating cosmologies upon time dependent hyperbolic space compactifications is called the space-like M2 (SM2) brane (for M theory)
or space-like D2 (SD2) brane (for string theory). Space-like branes \cite{Gutperle:2002ai} are topological defects localized on a space-like 
hypersurface and exist for a moment
in time. So, they are time dependent solutions of field theories or M/string theory with an isometry ISO($p+1$) $\times$ SO($d-p-2,1$) for an S$p$ brane
in $d$ space-time dimensions \cite{Chen:2002yq,Bhattacharya:2003sh}. The original motivation for constructing these solutions was to understand 
the time-dependent processes in field and M/string
theory \cite{Gutperle:2002ai,Sen:1999mg} and also to have a better understanding of the dS/CFT correspondence \cite{Strominger:2001pn}. The cosmological 
implication leading to four dimensional accelerated expansion
from these solutions has been elucidated in refs.\cite{Townsend:2003fx,Ohta:2003pu,Roy:2003nd,Ohta:2003ie}. 

The previous S2 brane solutions considered in the literature \cite{Ohta:2003pu,Roy:2003nd,Chen:2002yq,Bhattacharya:2003sh} were isotropic in the brane 
directions and so the four dimensional accelerating cosmologies 
obtained from these solutions were isotropic. In this paper we will construct an anisotropic SD2 brane solutions of type IIA string theory and try to see
whether similar four dimensional accelerating cosmologies can be obtained in all three spatial directions upon compactification. Another motivation
to look at the anisotropic SD2 brane solution is to see whether one can get a four dimensional Kasner-like \cite{Kasner:1921zz} solution from it upon 
compactification
where one can get expansions in all three spatial directions which is not possible in conventional Kasner solution from four dimensional vacuum Einstein
equation. The construction of anisotropic SD2 brane solution follows from the standard double Wick rotation \cite{Lu:2004ms} of the known anisotropic 
non-susy D2 brane 
solution \cite{Lu:2007bu} of type IIA string theory. This solution is characterized by five independent parameters. We then cast the solution in a 
suitable time-like coordinate
and is given in terms of a single harmonic function containing a characteristic time $t_0$. Next, we compactify the space-time on a six dimensional 
hyperbolic space with time dependent volume. The resultant metric when expressed in Einstein frame gives us a four dimensional FLRW type space-time with
three different scale factors in three spatial directions. We find that when $t \sim t_0$, we can get accelerating cosmologies in all three directions
when other parameters of the solution take some specific values. Although the expansions and the accelerations in all three directions are not the same 
but they do not 
differ drastically and the accelerations are all transient. However, when $t \ll t_0$, the resultant four dimensional metric takes a Kasner-like form
when the parameters characterizing the solution satisfy certain conditions.
But because of the presence of the dilaton as well as the volume scalar of the six dimensional hyperbolic space, all the Kasner exponents could be
positive definite, leading to expansions in all three spatial directions. However, the expansions in this case are decelerating. This
can be contrasted with the standard four dimensional Kasner space-time \cite{Kasner:1921zz} (obtained from the solution of vacuum Einstein equation) 
where expansions in all three directions are not possible.     
 
This paper is organized as follows. In the next section we give the construction of anisotropic SD2 brane solution from its time-like counterpart
and cast the solution in a coordinate system suitable for our purpose. In section 3, we obtain the anisotropic accelerating cosmologies from this 
solution upon compactification on six dimensional hyperbolic space of time dependent volume. In section 4, we show how a four dimesional Kasner-like 
geometry arises from this string theory solution, where all the Kasner exponents could be positive definite leading to expansions in all three spatial 
directions unlike the standard Kasner solution in four dimensions. Finally, we conclude in section 5.

\vspace{.5cm}

\noindent{\it 2. Anisotropic SD2 brane solutions} :
In this section we construct the anisotropic SD2 brane solution from the known anisotropic non-susy D2 brane solution of 
type IIA string theory. In \cite{Lu:2007bu}, we have constructed an anisotropic non-susy D$p$ brane solution and showed how it nicely interpolates
between a black D$p$ brane and a Kaluza-Klein ``bubble of nothing'' when some of the parameters of the solution are varied continuously
and interpreted this interpolation as closed string tachyon condensation. Here we make use of that solution and write the anisotropic
non-susy D2 brane solution in the following by putting $p=2$ in eq.(4) (we have replaced $\d_0$ by $\d_3$ for convenience) of the above mentioned 
reference \cite{Lu:2007bu},    
\bea\label{dp1}
ds^2 &=& F(r)^{\frac{3}{8}} \left(H(r)\tilde{H}(r)\right)^{\frac{2}{5}}\left(\frac{H(r)}{\tilde{H}(r)}\right)^{\frac{\delta_1}{4}
+\frac{\delta_2}{10}+\frac{\delta_3}{10}}\left(dr^2 + r^2 d\Omega_6^2\right)\nnu\\
& & + F(r)^{-\frac{5}{8}}\Big\{-\left(\frac{H(r)}{\tilde{H}(r)}\right)^{\frac{\delta_1}{4}+\frac{\delta_2}{2}+\frac{\delta_3}{2}}dt^2
+\left(\frac{H(r)}{\tilde{H}(r)}\right)^{-\frac{3\delta_1}{4}-\frac{3\delta_2}{2}+\frac{\delta_3}{2}}(dx^1)^2\nnu\\
& & + \left(\frac{H(r)}{\tilde{H}(r)}\right)^{-\frac{3\delta_1}{4}+\frac{\delta_2}{2}-\frac{3\delta_3}{2}}(dx^2)^2\Big\}\\
e^{2(\phi-\phi_0)} &=& F(r)^{\frac{1}{2}}\left(\frac{H(r)}{\tilde{H}(r)}\right)^{\delta_1-2\delta_2-2\delta_3}, \qquad\qquad F_{[6]} \,\,=\,\, 
\hat Q {\rm Vol}(\Omega_{6})
\eea
The metric in the above is given in the Einstein frame. The various functions appearing in the solution are defined as,
\bea\label{functions1}
F(r) &=& \left(\frac{H(r)}{\tilde{H}(r)}\right)^{\alpha} \cosh^2 \theta - \left(\frac{\tilde {H}(r)}{H(r)}\right)^{\beta} \sinh^2\theta\nnu\\
H(r) &=& 1 + \frac{\omega^5}{r^5},\qquad\qquad \tilde{H}(r)\,\,=\,\, 1 - \frac{\omega^5}{r^5}
\eea
Note that the solution has eight parameters $\alpha,\,\beta,\,\delta_1,\,\delta_2,\,\delta_3,\,\theta,\,\omega,$ and 
$\hat Q$. $\phi_0$ is the asymptotic value of the dilaton and $F_{[6]}$ is a six form and $\hat Q$ is the magnetic charge associated 
with the D2 brane. The solution becomes isotropic in the brane directions when $\d_1 = -2\d_2 = -2\d_3$. So, in that sense these
parameters can be called anisotropy parameters. Now for the consistency of the field equations the eight parameters of the solution
must satisfy the following relations \cite{Lu:2007bu},
\bea\label{relations1}
& & \alpha - \beta\,\,=\,\, -\frac{3}{2}\delta_1\nn
& & \frac{1}{2}\delta_1^2 + \frac{1}{2}\alpha(\alpha+\frac{3}{2}\delta_1)+\frac{2}{5}\delta_2\delta_3\,\,=\,\, \frac{6}{5}\left(1-\delta_2^2-\delta_3^2\right)\nn
& & \hat Q \,\,=\,\, 5 \omega^5(\alpha+\beta)\sinh2\theta
\eea
These three relations reduce the number of independent parameters from eight to five, which are $\omega$, 
$\theta$, and the anisotropy parameters $\delta_1,\,\delta_2$ and $\delta_3$. 
Using the second and the first relations in \eqref{relations1}, we can express $\a$ and $\b$ in terms of the other 
parameters as, 
\bea\label{alphabeta}
& & \alpha=-\frac{3}{4}\delta_1\pm \frac{1}{2}\sqrt{\frac{48}{5}(1-\delta_2^2-\delta_3^2)-\frac{7}{4}\delta_1^2-\frac{16}{5}\delta_2\delta_3}\nn
& & \beta=\frac{3}{4}\delta_1\pm \frac{1}{2}\sqrt{\frac{48}{5}(1-\delta_2^2-\delta_3^2)-\frac{7}{4}\delta_1^2-\frac{16}{5}\delta_2\delta_3}
\eea
The form of the harmonic function $\tilde{H}(r)$ in \eqref{functions1} indicates that there is a naked singularity of the solution at $r=\omega$ 
and therefore, the solution is well defined only for $r>\omega$. Now we apply the double Wick rotation \cite{Lu:2004ms} $r \to i\tau$, $t \to -ix^3$ to the 
solution \eqref{dp1} along with $\omega \to i\omega$, $\theta \to i\theta$ and $\theta_1 \to i\theta_1$, where $\theta_1$ is one of the angular coordinates 
of the sphere $\Omega_6$ of the transverse space. This operation gives us anisotropic space-like D2 brane from the anisotropic static non-susy D2 brane and 
the change in the angular 
coordinate converts spherical $\Omega_6$ to hyperbolic $H_6$. Thus the transformed solution is,
\bea\label{sdp1}
ds^2 &=& F(\tau)^{\frac{3}{8}} \left(H(\tau)\tilde{H}(\tau)\right)^{\frac{2}{5}}\left(\frac{H(\tau)}{\tilde{H}(\tau)}\right)^{\frac{\delta_1}{4}
+\frac{\delta_2}{10}+\frac{\delta_3}{10}}\left(-d\tau^2 + \tau^2 dH_6^2\right)\nn
& & + F(\tau)^{-\frac{5}{8}}\Big\{\left(\frac{H(\tau)}{\tilde{H}(\tau)}\right)^{\frac{\delta_1}{4}+\frac{\delta_2}{2}+\frac{\delta_3}{2}}(dx^3)^2
+\left(\frac{H(\tau)}{\tilde{H}(\tau)}\right)^{-\frac{3\delta_1}{4}-\frac{3\delta_2}{2}+\frac{\delta_3}{2}}(dx^1)^2\\
& & + \left(\frac{H(\tau)}{\tilde{H}(\tau)}\right)^{-\frac{3\delta_1}{4}+\frac{\delta_2}{2}-\frac{3\delta_3}{2}}(dx^2)^2\Big\}\nn
e^{2(\phi-\phi_0)} &=& F(\tau)^{\frac{1}{2}}\left(\frac{H(\tau)}{\tilde{H}(\tau)}\right)^{\delta_1-2\delta_2-2\delta_3}, \qquad\qquad F_{[6]} \,\,=\,\, \hat Q {\rm Vol}(H_{6})
\eea
The various functions associated with the solution are also changed under the above rotation and are given below,
\bea\label{functions2}
F(\tau) &=& \left(\frac{H(\tau)}{\tilde{H}(\tau)}\right)^{\alpha} \cos^2 \theta + \left(\frac{\tilde {H}(\tau)}{H(\tau)}\right)^{\beta} \sin^2\theta\nn
H(\tau) &=& 1 + \frac{\omega^5}{\tau^5},\qquad\qquad \tilde{H}(\tau)\,\,=\,\, 1 - \frac{\omega^5}{\tau^5}
\eea
Thus we see that the anisotropic static non-susy D2 brane has been converted to anisotropic time dependent or space-like D2 brane. For the former 
solution the radial coordinate $r$ was transverse to the D2 brane's world-volume, whereas, for the latter the timelike coordinate $\tau$ is 
transverse to the SD2 brane's 
world-volume. The metric of the transverse sphere $d\Omega_6^2$ has been converted to negative of the metric of the hyperbolic space $dH_6^2$. 
The hyperbolic functions\, $\sinh^2\theta$\, and \,$\cosh^2\theta$\, become \,$-\sin^2\theta$\, and \,$\cos^2\theta$\, respectively, therefore, 
the relative sign 
of the two terms of the function $F(\tau)$ has been flipped. But the form field remains unchanged with $\hat Q\rightarrow-\hat Q$. Thus the 
first two parameter 
relations in \eqref{relations1} remain the same, while the last relation has changed to $\hat Q=5\omega^5(\alpha+\beta)\sin2\theta$.  
Now for our purpose we will make a coordinate transformation from $\tau$ to $t$ given by,
\be\label{trans}
\tau \,\,=\,\, t\left(\frac{1+\sqrt{g(t)}}{2}\right)^{\frac{2}{5}}, \qquad {\rm where,}\qquad g(t)\,\, =\,\, 1+\frac{4\omega^5}{t^5} \equiv 1 + \frac{t_0^5}{t^5}
\ee
Under this coordinate change we have,
\bea\label{fnschange}
& & H(\tau) = 1 + \frac{\omega^5}{\tau^5} = \frac{2\sqrt{g(t)}}{1+\sqrt{g(t)}}, \qquad \tilde{H}(\tau) =  1 - \frac{\omega^5}{\tau^5} 
= \frac{2}{1+\sqrt{g(t)}},\nn
& & H(\tau)\tilde{H}(\tau) = \frac{4\sqrt{g(t)}}{(1+\sqrt{g(t)})^2}, \qquad \frac{H(\tau)}{\tilde{H}(\tau)} = \sqrt{g(t)},\nn
& & -d\tau^2 + \tau^2 dH_6^2 = g(t)^{\frac{1}{5}}\left(-\frac{dt^2}{g(t)} + t^2 dH_6^2\right)
\eea 
Using \eqref{fnschange} we can rewrite the anisotropic SD2 brane solution given in \eqref{sdp1} as follows,
\bea\label{sdp2}
ds^2 &=& F(t)^{\frac{3}{8}} g(t)^{\frac{\delta_1}{8}+\frac{\delta_2}{20}+\frac{\delta_3}{20}+\frac{1}{5}}\left(-\frac{dt^2}{g(t)} + 
t^2 dH_6^2\right)
+ F(t)^{-\frac{5}{8}}\Big[g(t)^{-\frac{3\delta_1}{8}-\frac{3\delta_2}{4}+\frac{\delta_3}{4}}(dx^1)^2\nn
&& +g(t)^{-\frac{3\delta_1}{8}+\frac{\delta_2}{4}-\frac{3\delta_3}{4}}(dx^2)^2+g(t)^{\frac{\delta_1}{8}+\frac{\delta_2}{4}+\frac{\delta_3}{4}}(dx^3)^2\Big]\nn
e^{2(\phi-\phi_0)} &=& F(t)^{\frac{1}{2}} g(t)^{\frac{\delta_1}{2}-\delta_2-\delta_3}, \qquad\qquad F_{[6]} \,\,=\,\, \hat{Q} {\rm Vol}(H_6)
\eea   
where $g(t)$ is as given in \eqref{trans} and $F(t)$ is given by,
\be\label{ft}
F(t) = g(t)^{\frac{\alpha}{2}} \cos^2\theta + g(t)^{-\frac{\beta}{2}} \sin^2\theta
\ee
It is important to note that in the new coordinate, the original singularity at $\tau=\omega$ has been 
shifted to $t=0$. Also note that as $t \gg t_0$, $g(t),\, F(t) \to 1$ and therefore, the solution reduces to flat space. In the next 
two sections we will impose the assumption $t \sim t_0$ and also $t \ll t_0$ into the solution \eqref{sdp2} to see how one can get 
accelerating cosmology in the first case and a Kasner-like cosmology in the second case in (3+1) dimensions upon compactification.

\vspace{.5cm}
\noindent{\it 3. Compactification and accelerating cosmology} : In this section we will compactify the anisotropic SD2 brane solution
given in \eqref{sdp2} on a six dimensional hyperbolic space of time dependent volume and write the resultant four dimensional metric 
in the Einstein frame\footnote{Here one might ask that since hyperbolic spaces are in general non-compact in what sense are we
compactifying the ten dimensional space on six dimensional hyperbolic space and studying the four dimensional cosmology? To address this question
we remark that it is quite well-known how to construct compact hyperbolic manifolds (CHM) from hyperbolic spaces and there is a vast
mathematical literature some of which are given in \cite{Kaloper:2000jb}. In short, the CHM's are obtained from $H_d$ (with $d\geq 2$), the
universal covering space of $d$ dimensional hyperbolic manifold by modding out by an appropriate freely acting discrete subgroup of the isometry group
SO$(1,d)$ of $H_d$. CHM's have many interesting properties and we refer the reader to some of the original literature given in \cite{Kaloper:2000jb} 
for details.}. This four dimensional metric will have the standard FLRW form whose cosmology we want to study.  
We rewrite the metric in \eqref{sdp2} in a four dimensional part and the transverse six dimensional part as,
\be\label{compct1}
ds^2 = ds_4^2+e^{2\psi}dH_6^2
\ee
where $\psi$ is the radion field and $e^{2\psi}=F(t)^{\frac{3}{8}} g(t)^{\frac{\delta_1}{8}+\frac{\delta_2}{20}+\frac{\delta_3}{20}+\frac{1}{5}}t^2$. The
four dimensional metric $ds_4^2$ is given as, 
\bea\label{4dmetric}
ds_4^2 &=& -F(t)^{\frac{3}{8}} g(t)^{\frac{\delta_1}{8}+\frac{\delta_2}{20}+\frac{\delta_3}{20}-\frac{4}{5}}dt^2
+ F(t)^{-\frac{5}{8}}\Big[g(t)^{-\frac{3\delta_1}{8}-\frac{3\delta_2}{4}+\frac{\delta_3}{4}}(dx^1)^2\nn
&&+g(t)^{-\frac{3\delta_1}{8}+\frac{\delta_2}{4}-\frac{3\delta_3}{4}}(dx^2)^2+g(t)^{\frac{\delta_1}{8}+\frac{\delta_2}{4}+\frac{\delta_3}{4}}(dx^3)^2\Big]
\eea
The compactified four dimensional metric \eqref{4dmetric} when expressed in Einstein frame takes the form \cite{Roy:2003nd}, 
\bea\label{einstein4d}
ds_{4E}^2 &=& e^{6\psi} ds_4^2\nn 
& = & -F(t)^{\frac{3}{2}}g(t)^{-{\frac{1}{5}}+{\frac{\delta_1}{2}}+{\frac{\delta_2}{5}}+{\frac{\delta_3}{5}}}t^6dt^2 + F(t)^{\half}g(t)^{{\frac{3}{5}}-{\frac{3\delta_2}{5}}+{\frac{2\delta_3}{5}}}t^6dx_1^2 \nn
&& + F(t)^{\half}g(t)^{{\frac{3}{5}}+{\frac{2\delta_2}{5}}-{\frac{3\delta_3}{5}}}t^6dx_2^2 + F(t)^{\half}g(t)^{{\frac{3}{5}}+{\frac{\delta_1}{2}}+{\frac{2\delta_2}{5}}+{\frac{2\delta_3}{5}}}t^6dx_3^2\nn
& = & -A(t)^2dt^2+\sum_{i=1}^3S_i(t)^2dx_i^2
\eea
where the various time-dependent coefficients are
\bea
 A(t) &=& F(t)^{\frac{3}{4}}g(t)^{-{\frac{1}{10}}+{\frac{\delta_1}{4}}+{\frac{\delta_2}{10}}+{\frac{\delta_3}{10}}}t^3, \quad\quad S_1(t)=F(t)^{\frac{1}{4}}g(t)^{{\frac{3}{10}}
-{\frac{3\delta_2}{10}}+{\frac{\delta_3}{5}}}t^3\nn
S_2(t) &= & F(t)^{\frac{1}{4}}g(t)^{{\frac{3}{10}}+{\frac{\delta_2}{5}}-{\frac{3\delta_3}{10}}}t^3, \quad\quad\qquad S_3(t)=F(t)^{\frac{1}{4}}g(t)^{{\frac{3}{10}}
+{\frac{\delta_1}{4}}+{\frac{\delta_2}{5}}+{\frac{\delta_3}{5}}}t^3
\eea
Note that in the compactified four dimensional space there are three fields, namely $g_{\mu\nu},\,\,\phi,\,\,\psi$. Now we perform another 
coordinate transformation
\be
d\eta^2=F(t)^{\frac{3}{2}}g(t)^{-{\frac{1}{5}}+{\frac{\delta_1}{2}}+{\frac{\delta_2}{5}}+{\frac{\delta_3}{5}}}t^6dt^2\,\,\,
\Rightarrow\,\,\eta= \int F(t)^{\frac{3}{4}}g(t)^{-{\frac{1}{10}}+{\frac{\delta_1}{4}}+{\frac{\delta_2}{10}}+{\frac{\delta_3}{10}}}t^3dt
\ee
and rewrite the Einstein frame metric $ds_{4E}^2$ in the standard flat FLRW form as
\be\label{flrw4}
ds_{4E}^2 = - d\eta^2 + s_i^2(\eta)\sum_{i=1}^{3} (dx^i)^2
\ee
with $\eta$ being the canonical time and the scale factor $s_i(\eta)\equiv S_i(t)$. Note that since $s_i(\eta)$ are different for each $i$, the cosmology here
will be anisotropic. Now because of the complicated relation between $t$ and $\eta$ let us define \cite{ourpaper}
\bea\label{mandn}
m_i(t) & \equiv & \frac{d\ln S_i(t)}{dt}\nn
n_i(t) & \equiv &\left[\frac{d^2}{dt^2}\ln(S_i(t))+\frac{d}{ dt}\ln(S_i(t))
\frac{d}{dt}\ln\left(\frac{S_i(t)}{A(t)}\right)\right]
\eea
and with these one can easily see that $m_i(t) > 0$ implies that $ds_i(\eta)/d\eta > 0$, amounting to expansion of our universe, and similarly,
$n_i(t) > 0$ implies that $d^2 s_i(\eta)/d\eta^2 > 0$, amounting to acceleration of our universe.
Therefore, from \eqref{mandn} it is clear that 
in the four-dimensional spacetime \eqref{einstein4d} we get an accelerated expansion in the $i$-th coordinate direction only if the 
parameters $m_i(t)$ and $n_i(t)$ are simultaneously positive in that direction. It can be checked that for $t \ll t_0$, accelerating
expansion is not possible at all in any direction. However, it is possible only if $t \sim t_0$. In this case the first term in
the harmonic function $g(t)$ given in \eqref{trans} is of the same order as the second. The other parameters of the solution, namely, 
$\d_1$, $\d_2$ and $\d_3$ can not be totally arbitrary. From Eq.\eqref{alphabeta}, we see that the reality of $\a$
and $\b$ imposes some restriction on the value of these three anisotropy parameters. Also it can be checked that by changing the value of 
$\theta$ does not change the cosmological behavior of the solution very much. Thus we have chosen some typical values of these parameters
(as given in the Figure) and plotted the functions $m_i(t)$ and $n_i(t)$ in Figure 1, to show that it is indeed possible to have accelerating 
expansions in all three directions.       
\begin{figure}[ht]
\begin{center}
\subfloat[along $x_1$ direction]{\includegraphics[width=0.33\textwidth]{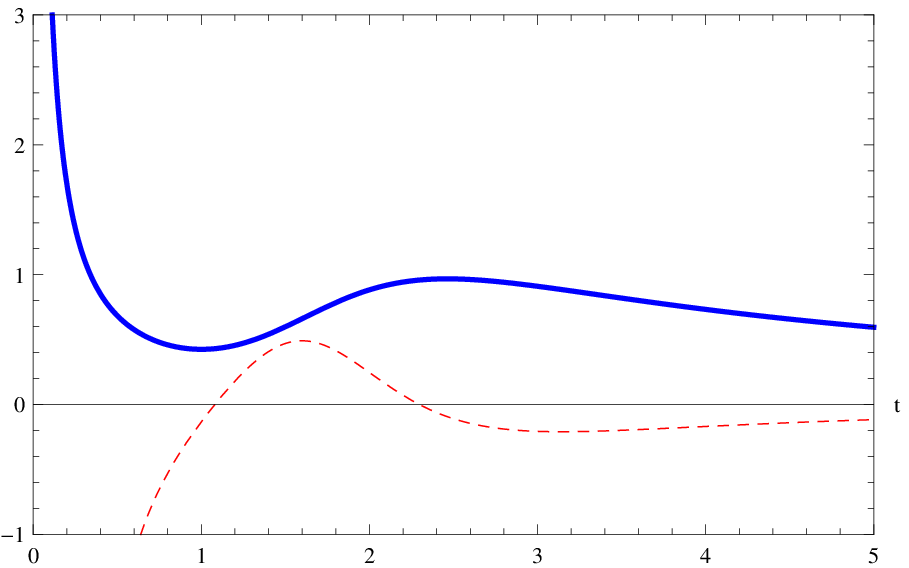}\label{pic1}}
\subfloat[along $x_2$ direction]{\includegraphics[width=0.33\textwidth]{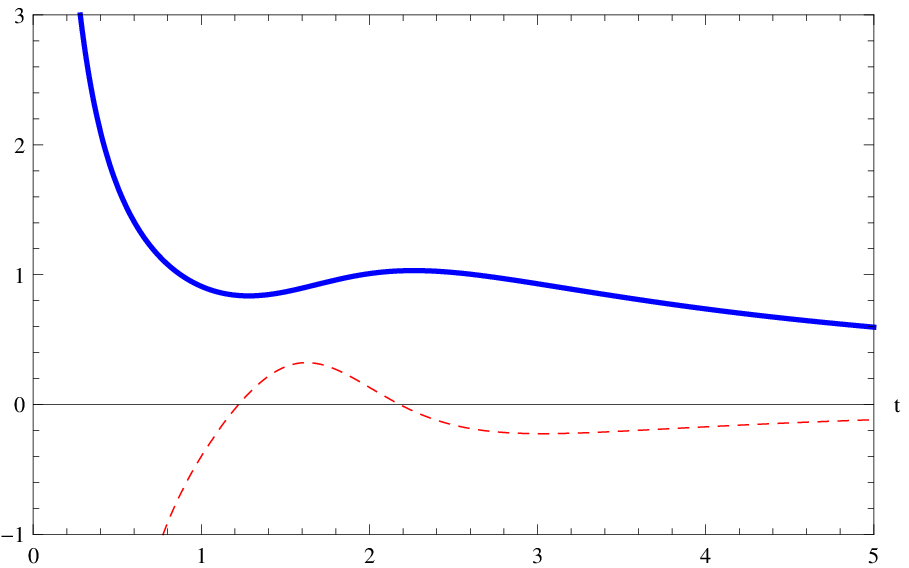}\label{pic2}}\subfloat[along $x_3$ direction]
{\includegraphics[width=0.33\textwidth]{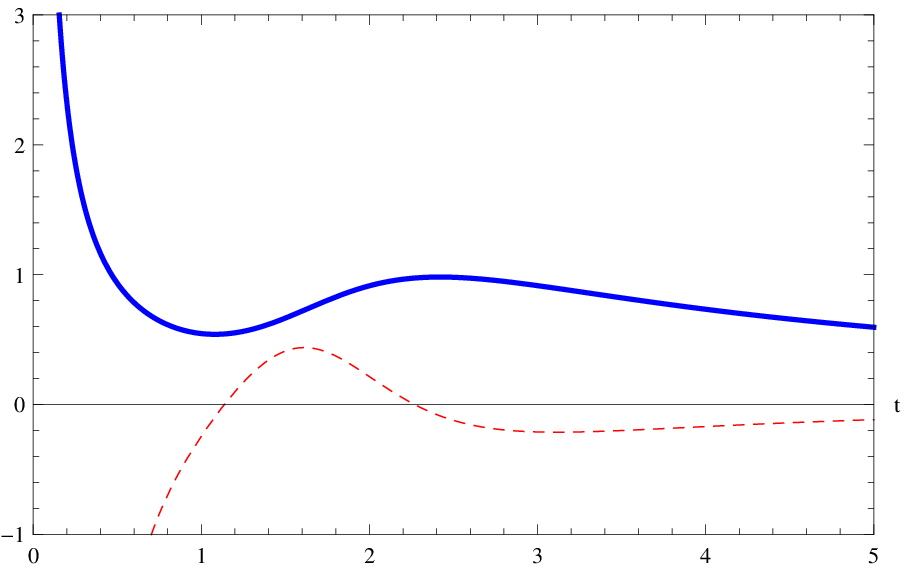}\label{pic3}}
\end{center}
\caption[\textwidth]{The plot of $m(t)$(solid blue line) and $n(t)$(dashed red line) in different spatial coordinate directions at 
$\theta=\pi/6,\,\delta_1=-0.5,\,\delta_2=0.2,\,\delta_3=0.4$ and $t_0=2.0$.}
\end{figure}
We notice as shown in (a), (b) and (c) in Figure 1, we always get expanding universe (given by the solid
blue line) in all three directions, but the expansion is accelerating only for a short period of time, i.e., the acceleration is transient 
(given by the dotted red line). Also note that since $m_i(t)$ and $n_i(t)$ are different for different $i$, the cosmology is anisotropic, 
however, the anisotropy is not too much.

To understand the accelerating expansion, we can write down the four dimensional compactified action from the original ten dimensional one 
and obtain the form of the potential of the dilaton and the radion field \cite{Roy:2003nd}. The ten dimensional action has the form,
\be\label{10daction}
S = \int d^{10}x \sqrt{-g}\left[R - \half (\partial\phi)^2 - \frac{1}{2 \cdot 6!} e^{-\phi/2} F_{[6]}^2\right]
\ee
Reducing the action on a six dimensional hyperbolic space $H_6$, the four dimensional action we get\footnote{Here reduction on the
hyperbolic space $H_6$ to obtain the four dimensional action is done in the sense decribed in footnote 3. This has also been done
in the references \cite{Garriga:2000cv,Emparan}.} \cite{ourpaper,Garriga:2000cv,Emparan}
\be\label{4daction}
S_4 = \int d^4 x \sqrt{-g_{4E}} \left[R_{4E} - \half (\partial \phi)^2 - 24 (\partial \psi)^2 - V(\phi,\psi)\right]
\ee
where,
\be\label{potential}
V(\phi,\psi) = \frac{\hat{Q}^2}{2} e^{-\frac{\phi}{2} - 18\psi} + 30 e^{-8\psi}.
\ee
Here $\hat{Q}$ is the magnetic charge of the D2 brane given in \eqref{dp1}. Note that because of the hyperbolic space compactification the 
potential is always positive irrespective of the charge and therefore there is always a possibility that the system will be driven 
to an accelerating phase \cite{Emparan}. 

\vspace{.5cm}

\noindent{\it 4. Compactification and Kasner-like solution} : 
In this section we will show how  a four dimensional Kasner-like cosmological solution follows from the anisotropic SD2 brane solution
upon six dimensional hyperbolic space compactification discussed in the previous section. The compactified action expressed in
Einstein frame is given in \eqref{einstein4d}. We take this four dimensional metric and express it at early times, $t  \ll t_0$. 
In this case the function $g(t)$ can be approximated as, 
\be\label{gtau} 
g(t)= 1 + \frac{t_0^5}{t^5} \approx \frac{t_0^5}{t^5} \sim t^{-5},
\ee
Also since we want to express the metric components in \eqref{einstein4d} as some powers of $t$, we note from the form of $F(t)$ in
\eqref{ft} that this can be done (assuming $\a>0$ without any loss of generality) in three ways as follows. (a) Put $\theta=0$, with $\a$, $\b$ 
as given in \eqref{alphabeta}, (b) put $\a=-\b = -(3/4)\d_1$, with $\theta$ arbitrary and (c) both $\a > 0$, $\b > 0$, with $\theta$ arbitrary.
There is another possibility with $\theta = \pi/2$ and $\b < 0$, but this case can be seen to be equivalent to case (a).
Note that for case (a) and (b) we have $\hat{Q}=0$ (since $\hat{Q}=5\omega^5(\a+\b)\sin2\theta$), however, for case (c) $\hat{Q}$ is non-zero
and the non-susy brane is magnetically charged. 
In either case (a) or (b) we have  
\be\label{Ft}
F(t) \sim t^{-\frac{5\alpha}{2}} 
\ee
In the above we have absorbed $t_0$ in $t$. But for case (c) $F(t)$ has an additional $\cos^2\theta$ factor which can be absorbed in $t$ as well
as in $x^{1,2,3}$. Thus in all cases $F(t)$ has the form as given in \eqref{Ft}. 
So,
in this near region, the space-time metric \eqref{einstein4d}, the dilaton and the radion fields take the forms,
\bea\label{kasner1}
ds^2 &=& -\,t^{2\left(\frac{7}{2}-\frac{15\alpha}{8}-\frac{5\delta_1}{4}-\frac{\delta_2}{2}-\frac{\delta_3}{2}\right)} dt^2 + t^{2\left(\frac{3}{2}-\frac{5\alpha}{8}+\frac{3\delta_2}{2}
-\delta_3\right)} (dx^1)^2\nn
& &+\,t^{2\left(\frac{3}{2}-\frac{5\alpha}{8}-\delta_2+\frac{3\delta_3}{2}\right)} (dx^2)^2+t^{2\left(\frac{3}{2}-\frac{5\alpha}{8}-\frac{5\delta_1}{4}-\delta_2-\delta_3\right)} (dx^3)^2\nn 
e^{2(\phi-\phi_0)} &=& t^{2\left(-\frac{5\alpha}{8}-\frac{5\delta_1}{4}+\frac{5\delta_2}{2}+\frac{5\delta_3}{2}\right)}, \qquad
e^{2\psi} = t^{2\left({\half}-\frac{15}{32}\alpha-\frac{5}{16}\delta_1-\frac{\delta_2}{8}-\frac{\delta_3}{8}\right)}
\eea
Now since we are taking $t\ll 1$ here, we have to be careful about the validity of the gravity solution. The gravity solution will
be valid as long as the dilaton remains small and the curvature of the transverse space in string units also remains small. These two
conditions impose certain restrictions on the parameters of the solution and they are given as,
\bea\label{constraints}
& & 5 \a + 5\d_1 - 4\d_2 - 4\d_3 > 4\nn
& & \a + 2\d_1 - 4\d_2 - 4\d_3 \leq 0
\eea
where $\a$ is as given in \eqref{alphabeta}. Furthermore, the reality of $\a$ also restricts the parameters as 
\be\label{cons}
\frac{35}{4}\d_1^2 + 48 \d_2^2 + 48 \d_3^2 + 16 \d_2\d_3 \leq 48
\ee
We have checked numerically that all these three conditions can be satisfied simultaneously for certain range of values of the parameters
$\d_1$, $\d_2$ and $\d_3$ and only for those values we have a valid gravity solution \eqref{kasner1}. We would like to remark here that
the validity of the supergravity solution also requires that we cannot take $t$ arbitrarily close to zero as we are considering $t\ll 1$.
In fact $t$ has to be much larger than the string scale if the supergravity solution remains valid. This can be seen if we calculate $\dot{\phi}^2$,
$\dot{\psi}^2$ and also the scalar curvature with the solution given in \eqref{kasner1}. All of these terms come out to be proportional to
$1/t^2$ and so, when $t \ll 1$, they can become very large invalidating the supergravity solution and stringy corrections must be included.
To avoid this we require $\sqrt{\a'} \ll t \ll t_0$ or in terms of scaled $t$, we must have $\sqrt{\a'}/t_0 \ll t \ll 1$.   
Now, keeping those restrictions in mind, we can rewrite the solution in terms of canonical time 
$\eta \equiv \frac{8t^{\frac{9}{2}-\frac{15\alpha}{8}-\frac{5\delta_1}{4}-\frac{\delta_2}{2}-\frac{\delta_3}{2}}}
{36-15\alpha-10\delta_1-4\delta_2-4\delta_3}$ as,
\bea\label{kasner2}
ds^2 &=& -d\eta^2+\eta^{2p_1}(dx^1)^2+\eta^{2p_2}(dx^2)^2+\eta^{2p_3}(dx^3)^2\nn 
e^{2(\phi-\phi_0)} &=& C(\d_1,\d_2,\d_3)\,\eta^{2\gamma_\phi}\qquad\qquad e^{2\psi}= D(\d_1,\d_2,\d_3)\eta^{2\gamma_\psi}
\eea
Note that in writing the metric in \eqref{kasner2} we have rescaled the coordinates $x^1$, $x^2$ and $x^3$ by some constant factors involving
the parameters $\d_1$, $\d_2$, $\d_3$. Also in the dilaton and the radion field $C$ and $D$ are constants involving these
paramaters whose explicit form will not be important. It can be easily checked in the defining relation of $\eta$, that the coefficient
in front of $t$ is always positive definite and that also ensures that as $t \to 0$, $\eta \to 0$.   
The Kasner exponents $p_1$, $p_2$ and $p_3$ in the metric and $\gamma_\phi$,
$\gamma_\psi$ are defined as,
\bea\label{kasnercoeff}
& & p_1=\frac{12-5\alpha+12\delta_2-8\delta_3}{36-15\alpha-10\delta_1-4\delta_2-4\delta_3}\nn
& & p_2=\frac{12-5\alpha-8\delta_2+12\delta_3}{36-15\alpha-10\delta_1-4\delta_2-4\delta_3}\nn
& & p_3=\frac{12-5\alpha-10\delta_1-8\delta_2-8\delta_3}{36-15\alpha-10\delta_1-4\delta_2-4\delta_3}\nn
& & \gamma_\phi=\frac{-5\alpha-10\delta_1+20\delta_2+20\delta_3}{36-15\alpha-10\delta_1-4\delta_2-4\delta_3}\nn
& & \gamma_\psi=\frac{1}{4}\frac{16-15\alpha-10\delta_1-4\delta_2-4\delta_3}{36-15\alpha-10\delta_1-4\delta_2-4\delta_3}
\eea
Now since this a solution to the compactified four dimensional action given in \eqref{4daction}, it must satisfy the equations of motion. 
The Einstein equation, the dilaton and radion equations following from \eqref{4daction} have the forms,
\bea\label{einsteineq}
& & R_{\mu\nu,E}-{\half}\partial_\mu\phi\partial_\nu\phi-24\partial_\mu\psi\partial_\nu\psi=0\nn
& & \frac{1}{\sqrt{-g_E}}\partial_\mu\left(\sqrt{-g_E}g^{\mu\nu}_E \partial_\nu\phi\right) = 0, \qquad 
\frac{1}{\sqrt{-g_E}}\partial_\mu\left(\sqrt{-g_E}g^{\mu\nu}_E \partial_\nu\psi\right) = 0
\eea
here $\mu,\,\nu$ run over $(1+3)$-dimensional space-time. Note that since we have $t \ll 1$, the potential in \eqref{potential} is trivial
(the first term is zero even when $\hat{Q} \neq 0$ because the exponential factor effectively goes to zero due to the relations given in \eqref{constraints}
and similarly the exponential in the second term also effectively goes to zero because of \eqref{constraints}).   
Substituting the above solution \eqref{kasner2} in \eqref{einsteineq}, we get two conditions 
\be\label{pcond1}
p_1+p_2+p_3=1, \qquad {\rm and} \qquad p_1^2+p_2^2+p_3^2=1-\half \gamma_\phi^2-24\gamma_\psi^2
\ee
The first condition of \eqref{pcond1} can be seen to be satisfied trivially from \eqref{kasnercoeff}. On the other hand when we
substitute the parameter values from \eqref{kasnercoeff} to the second condition of \eqref{pcond1}, we find that it gives the same 
parametric relation as the second relation of \eqref{relations1} verifying the consistency of the solution. This therefore shows how 
one can get a four dimensional Kasner-like
solution from the ten dimensional anisotropic SD2 brane solution by six dimensional hyperbolic space compactification. It is well-known that
the standard Kasner solution \cite{Kasner:1921zz} obtained as the solution of vacuum Einstein equation, does not lead to expansions in all spatial directions.
The reason is that in standard Kasner cosmology the Kasner exponents satisfy $p_1+p_2+p_3=1$ and $p_1^2+p_2^2+p_3^2=1$. Since these two
conditions can not be satisfied together when $p_i$'s are all positive, the expansions can not occur in all the directions. However, for the 
four-dimensional Kasner cosmology we obtained from string theory
solutions, the parameters $p_i$'s can all be positive definite because the second condition here \eqref{pcond1} is different. This is the 
essentially the reason that we can have expansions in all the directions, but, it can be easily checked that the expansions are decelerating.      

\vspace{.5cm}

\noindent{\it 5. Conclusion} : To summarize, in this paper we have constructed an anisotropic SD2 brane solution starting from
an anisotropic non-susy D2 brane solution of type IIA string theory by the standard trick of double Wick rotation. We wanted to see whether
it is possible to generate accelerating cosmologies in all the directions which is known for the isotropic SD2 brane solution upon
compactification on six dimensional hyperbolic space of time dependent volume. Indeed we found that when the resultant four dimensional
metric is expressed in Einstein frame there are some windows of the parameters
of the solution where one can get accelerating cosmologies in all the directions and is discussed in section 3. Here both the expansions
and the accelerations we found are anisotropic. But, in order to get accelerating expansions we noted that the anisotropy can not
be too drastic in three different directions. We also noted that accelerations are possible only for $t \sim t_0$, where $t_0$ is some
characteristic time given as one of the parameters of the solution. Next, we looked at the four dimensional metric at early times,
i.e., for $t \ll t_0$ and found that in a suitable coordinate and under certain conditions on the parameters of the solution, it can be expressed 
in a standard four dimensional Kasner-like form.
But unlike in the standard Kasner cosmology, where expansions in all three directions are not possible, here we can get expansions
in all the three directions. The reason is that in this case the relations among the Kasner exponents get modified due to the presence
of the dilaton and the radion field. It would be interesting to see what effect (such modification to Kasner solution at early time) 
does it have on the cosmological singularities \cite{Engelhardt:2014mea,Chatterjee:2016bhj}.

\vspace{.5cm}


\begin{thebibliography}{99}

\bibitem{Townsend:2003fx} 
  P.~K.~Townsend and M.~N.~R.~Wohlfarth,
  ``Accelerating cosmologies from compactification,''
  Phys.\ Rev.\ Lett.\  {\bf 91}, 061302 (2003)
  [hep-th/0303097].


\bibitem{Kaloper:2000jb} 
  N.~Kaloper, J.~March-Russell, G.~D.~Starkman and M.~Trodden,
  ``Compact hyperbolic extra dimensions: Branes, Kaluza-Klein modes and cosmology,''
  Phys.\ Rev.\ Lett.\  {\bf 85}, 928 (2000)
  [hep-ph/0002001];
  G.~D.~Starkman, D.~Stojkovic and M.~Trodden,
  ``Large extra dimensions and cosmological problems,''
  Phys.\ Rev.\ D {\bf 63}, 103511 (2001)
  [hep-th/0012226];
  G.~D.~Starkman, D.~Stojkovic and M.~Trodden,
  ``Homogeneity, flatness and 'large' extra dimensions,''
  Phys.\ Rev.\ Lett.\  {\bf 87}, 231303 (2001)
  [hep-th/0106143].

\bibitem{Gibbons:1985}
G. W. Gibbons, ``Aspects of supergravity theories,'' in {\it Supersymmetry, Supergravity and
Related Topics}, Eds. F. de Aguila, J. A. de Azcarraga and L. Ibanez, p.346 (World Scientific,
Singapore, 1985).


\bibitem{Maldacena:2000mw} 
  J.~M.~Maldacena and C.~Nunez,
  ``Supergravity description of field theories on curved manifolds and a no go theorem,''
  Int.\ J.\ Mod.\ Phys.\ A {\bf 16}, 822 (2001)
  [hep-th/0007018].

\bibitem{Ohta:2003pu} 
  N.~Ohta,
  ``Accelerating cosmologies from S-branes,''
  Phys.\ Rev.\ Lett.\  {\bf 91}, 061303 (2003)
  [hep-th/0303238].

\bibitem{Roy:2003nd} 
  S.~Roy,
  ``Accelerating cosmologies from M / string theory compactifications,''
  Phys.\ Lett.\ B {\bf 567}, 322 (2003)
  [hep-th/0304084].

\bibitem{Gutperle:2002ai} 
  M.~Gutperle and A.~Strominger,
  ``Space - like branes,''
  JHEP {\bf 0204}, 018 (2002)
  [hep-th/0202210];
  A.~Maloney, A.~Strominger and X.~Yin,
  ``S-brane thermodynamics,''
  JHEP {\bf 0310}, 048 (2003)
  [hep-th/0302146].

\bibitem{Chen:2002yq} 
  C.~-M.~Chen, D.~V.~Gal'tsov and M.~Gutperle,
  ``S brane solutions in supergravity theories,''
  Phys.\ Rev.\ D {\bf 66}, 024043 (2002)
  [hep-th/0204071];
  M.~Kruczenski, R.~C.~Myers and A.~W.~Peet,
  ``Supergravity S-branes,''
  JHEP {\bf 0205}, 039 (2002)
  [hep-th/0204144];
  S.~Roy,
  ``On supergravity solutions of space - like Dp-branes,''
  JHEP {\bf 0208}, 025 (2002)
  [hep-th/0205198].

\bibitem{Bhattacharya:2003sh} 
  S.~Bhattacharya and S.~Roy,
  ``Time dependent supergravity solutions in arbitrary dimensions,''
  JHEP {\bf 0312}, 015 (2003)
  [hep-th/0309202].

\bibitem{Sen:1999mg} 
  A.~Sen,
  ``NonBPS states and Branes in string theory,''
  hep-th/9904207;
  A.~Sen,
  ``Rolling tachyon,''
  JHEP {\bf 0204}, 048 (2002)
  [hep-th/0203211].

\bibitem{Strominger:2001pn} 
  A.~Strominger,
  ``The dS / CFT correspondence,''
  JHEP {\bf 0110}, 034 (2001)
  [hep-th/0106113];
  M.~Spradlin, A.~Strominger and A.~Volovich,
  ``Les Houches lectures on de Sitter space,''
  hep-th/0110007.

\bibitem{Ohta:2003ie} 
  N.~Ohta,
  ``A Study of accelerating cosmologies from superstring / M theories,''
  Prog.\ Theor.\ Phys.\  {\bf 110}, 269 (2003)
  [hep-th/0304172].
  R.~Emparan and J.~Garriga,
  ``A Note on accelerating cosmologies from compactifications and S branes,''
  JHEP {\bf 0305}, 028 (2003)
  [hep-th/0304124];
  C.~-M.~Chen, P.~-M.~Ho, I.~P.~Neupane, N.~Ohta and J.~E.~Wang,
  ``Hyperbolic space cosmologies,''
  JHEP {\bf 0310}, 058 (2003)
  [hep-th/0306291];
  S.~B.~Giddings and R.~C.~Myers,
  ``Spontaneous decompactification,''
  Phys.\ Rev.\ D {\bf 70}, 046005 (2004)
  [hep-th/0404220];
S. Roy and H. Singh, 
``Space-like branes, accelerating cosmologies and the near `horizon' limit,''
JHEP {\bf 0608}, 024 (2006) [hep-th/0606041];
  K.~Nayek and S.~Roy,
``Space-like D$p$ branes: accelerating cosmologies versus conformally de Sitter space-time,''
  JHEP {\bf 1502}, 021 (2015)
  [arXiv:1411.2444[hep-th]]

\bibitem{Kasner:1921zz} 
  E.~Kasner,
  ``Geometrical theorems on Einstein's cosmological equations,''
  Am.\ J.\ Math.\  {\bf 43}, 217 (1921).

\bibitem{Lu:2004ms} 
  J.~X.~Lu and S.~Roy,
  ``Static, non-SUSY p-branes in diverse dimensions,''
  JHEP {\bf 0502}, 001 (2005)
  [hep-th/0408242].

\bibitem{Lu:2007bu} 
  J.~X.~Lu, S.~Roy, Z.~L.~Wang and R.~J.~Wu,
  ``Intersecting non-SUSY branes and closed string tachyon condensation,''
  Nucl.\ Phys.\ B {\bf 813}, 259 (2009)
  [arXiv:0710.5233 [hep-th]].


\bibitem{ourpaper}
See, for example, the last two references of [12]. 

\bibitem{Garriga:2000cv} 
  J.~Garriga and A.~Vilenkin,
  ``Solutions to the cosmological constant problems,''
  Phys.\ Rev.\ D {\bf 64}, 023517 (2001)
  doi:10.1103/PhysRevD.64.023517
  [hep-th/0011262].
 
\bibitem{Emparan}
See the second reference in [12]. 

\bibitem{Engelhardt:2014mea} 
  N.~Engelhardt, T.~Hertog and G.~T.~Horowitz,
  ``Holographic Signatures of Cosmological Singularities,''
  Phys.\ Rev.\ Lett.\  {\bf 113}, 121602 (2014)
  [arXiv:1404.2309 [hep-th]];
  N.~Engelhardt, T.~Hertog and G.~T.~Horowitz,
  ``Further Holographic Investigations of Big Bang Singularities,''
  JHEP {\bf 1507}, 044 (2015)
  [arXiv:1503.08838 [hep-th]];

\bibitem{Chatterjee:2016bhj} 
  S.~Chatterjee, S.~P.~Chowdhury, S.~Mukherji and Y.~K.~Srivastava,
  ``Non-vacuum AdS cosmology and comments on gauge theory correlator,''
  arXiv:1608.08401 [hep-th].







\end{thebibliography}
\end{document}